\documentclass[12pt]{article}
\input{epsf.tex}
\begin{document}
\begin{center}
{\bf POSSIBILITIES TO VERIFY THE LEVEL DENSITY AND RADIATIVE
STRENGTH FUNCTIONS, EXTRACTED FROM THE TWO-STEP $\gamma$-CASCADE INTENSITIES}
\end{center}\begin{center}
 {\bf A.M. Sukhovoj, V.A. Khitrov }\\
 {\it Frank Laboratory of Neutron Physics, Joint Institute for Nuclear
Research \\ 141980 Dubna, Russia}\\
{ \bf Pham Dinh Khang}\\
{\it National University of Hanoi, Vietnam }\\
{ \bf Vuong Huu Tan,  Nguyen Xuan Hai}\\
{\it Vietnam Atomic Energy Commission, Vietnam }\\
\end{center}
\section{Introduction}

The direct determination of density $\rho$ of excitation levels
(number of levels of nucleus in the unit interval of excitation energy)
for the larger part of stable and long-life radioactive target nuclei is
impossible.
This assertion relates also to the radiative strength functions
\begin{equation} k=\Gamma_{\lambda i}/(E_{\gamma}^3\times
A^{2/3}\times D_{\lambda})
\end{equation}
exciting their primary dipole electrical and magnetic gamma-transitions of
level of nucleus decaying the excited in the nuclear reaction.
This circumstance is uniquely determined by the fact, that
$D_{\lambda}$is 
much less than the resolution of the existing spectrometers of gamma-rays
and charged particles or neutrons. Extraction of the parameters of nucleus
in question in this situation can be executed by only their fitting to the
optimum values,
reproducing the  spectra and cross section with
the minimum standard deviation measured in the nuclear reactions.

This inverse problem of mathematical analysis by its nature is principally
ambiguous. Moreover, systems of equations, connecting the number of excited
levels and probability of the emission of nuclear products are  usually 
assigned
within the framework of some ideas about the mechanism of nuclear reaction and
factors determining the dynamics of the studied process.

Thus, for example, the description of the cascade gamma-decay of neutron
resonance, is impossible at present without the introduction of some
a priori ideas. In particular, within the framework of 
the ideas about this process  following potential possibilities are not taken
into consideration:

the presence of the strong dependence of neutron widths
$\Gamma_n^0$ on the structure of the
wave function of resonances (and of the excessive error in determination of their
density by the time-of-flight method),

the analogous dependence of the partial radiative widths of primary gamma-transitions
on the structure of the level excited by them, evidently overstepping the limits of
the expected Porter- Thomas fluctuations;

The possibility of existence of the enumerated effects and their sufficiently
their significant influence on the process of cascade gamma-decay directly
follows from the results [1] of model approximation of the level density, extracted from
the reaction $(n, 2\gamma)$ and the comparison [2,3] of the average values of the
sums of radiative strength functions with their models [4,5] most often used in
practice.
Thus, the break of sequential Cooper pair [1] with the excitation energy in the
region of the neutron-binding energy can change values of one (two) quasi-particle
components in the wave function and thus - change [6] values of $\Gamma_n^0$.
This possibility directly follows from the results, presented in [1].
Whether this possibility is realizable in principle, to what degree the process of
fragmentation of nuclear states mixes up components of different types in the
wave functions of the levels in the region $E_{ex} \geq B_n$
and at the noticeably higher excitation energies - neither the experiment, nor the theory
can answer this at the present.

In particular it is not possible to obtain the realistic estimation of the
part of the unobservable levels, which according to the values $J^{\pi}$ could be
excited them as s-resonances.
This problem is very essential, since the density of neutron resonances in practice
in any experiment to determine this parameter for the excitation lower than $B_n$ is used
to standardize its relative values.
As a consequence of the above mentioned facts, the measured experimentally in different
procedures [2,3,7,8] level densities and the radiative strength functions of primary
gamma-transitions can have an unknown systematic error, the value of which directly
depends on a systematic error in the conventional values
$D_{\lambda}$ of the spacing between
the neutron resonances.
And the obtained ideas about these values and properties of nucleus can be erroneous to
a greater or lesser extent.
However, if we add the fundamental incompatibility of the data about the level density
between the results of applying the procedures [7,8] on the one hand,
and [2,3] on the other hand, than the need for a maximally possible
verification of $\rho$ and $k$ determined from indirect experiments becomes obvious.

\section{Possibilities and the specific character of the verification of the experimentally
determined values of $\rho$ and 
 $k$}

The verification of the indicated parameters of nucleus can be partially
executed by the calculation of total gamma-spectra for different sets of
$\rho$ and $k$ with their subsequent comparison with the experiment.
This calculation was carried out by different authors repeatedly [9,10], but,
as a rule, without taken into account of:

the nonconformity of the model assigned ones and real values of $\rho$ or $k$
if to determine one of these values purely model presentations
about another value  are used;

the specific character of the transfer of errors $\delta \rho$ and $\delta k$
to an error of the calculated spectrum;

all aspects of the influence of the structure of the excited levels of nucleus
on the parameters $\rho$ and $k$.

All these problems become apparent to the full during the calculation of 
the gamma-ray spectra of the radiative capture of thermal
neutrons, measured, for example, by Groshev [11], with the use of $\rho$ and $k$, 
determined from the gamma-ray intensities
in the procedures [8] or [2,3]. The major part of experimental data on the total
gamma-ray spectra of the capture not only of thermal but also fast neutrons was
used to verify such data earlier [12]. At present, 
maximally reliable data on to the level density and the radiative
strength functions of primary and secondary gamma-transitions are acquired [2,3] for the
isotopes $^{183,184,185,187}$W.
They describe precise not only the intensities of two-step cascades and the
total radiative widths of the compound-states of these nuclei, but also, to the
maximum degree - the cascade population of their levels up to the excitation
energy 3-4 MeV. Noticeably less reliable (because of low efficiency of the used
spectrometer) 
Values $\rho$ and $k$ are determined also for the isotopes
$^{163,164,165}$Dy [2,3,13].

The corresponding total gamma-ray spectra are determined by the enumerated
isotopes for 88\% in Dy and practically for 100\% - in W.
Due to the absence of data on the level density and radiative strength functions
extracted from the cascade intensities for the compound nucleus of $^{162}$Dy,
the missing 12\% are completed by the total gamma-spectrum
of $^{164}$Dy.
This is done on the basis of the observation of insubstantial divergence in the
calculated total gamma-spectra of odd isotopes and extrapolation of this fact
to the even isotopes of dysprosium.

\section{The comparison of calculation and experiment}

It is most expedient to perform
the comparison of total gamma-spectra for the spectrum corresponding to the product
of the gamma-quantum
intensity on their energy.
The condition $\sum I_\gamma E_\gamma =B_n$ ensures the maximally precise normalization
intensities of the observed gamma-transitionson the average and the
presence of errors of different sign - for different values of gamma-quantum energies.
For the spectrum determined in [11] during the capture of
thermal neutrons in the target of natural isotopic composition the normalization
is cared out for the sum of
neutron-binding energy the weighted on the part of captures in all the isotopes.
For Dy and W it is equal 6104 and 5770 keV, respectively, and is determined
by the target isotopes $^{164}$Dy and $^{186}$W dominating in the capture .

The sums of the level densities of both parities and spins 1/2 and 3/2 for these
compound nuclei are given in Fig. 1;  the sums of the radiative strength functions
of primary gamma-transitions with the coefficients of an increase in the radiative
strength functions of secondary transitiones are given in Fig. 2 respectively.
The calculated total gamma-spectra of the capture of thermal neutrons in the
SAMPLE of natural isotopic composition are compared with the experiment in Fig. 3.
The real resolution of the Groshev's experiment
was considered at their calculation. Also in the distributions of the
cascade intensities determined in the experiment with the threshold of the registration of
520 keV the cascades are added
corresponding in the observed  intensity to the strongest  primary gamma-transitions to the
ground and one or two low-lying levels.

This procedure changes very weakly the parameters
of sets $\rho$ and $k(E1)+k(M1)$, of those determined
accordingly [2,3],  but it allow one to exclude the high-energy part
of the total gamma-spectra with their maximum fluctuations from the comparison.
Practically, this leads to reduction of the area of low-energy ($E_\gamma < 5$ MeV)
part of the area of the calculated spectrum less, than 5\%.
An error in the determination of total gamma-spectra can be estimated from the
comparison of the intensities of the same transitions with the contemporary data [14].
It cannot be less than 10-20\$.

The results of the comparison of the spectra, calculated for different functional dependencies
of level density and sums of the strength functions of dipole gamma-transitions
with the experiment, quite
unambiguously lead to the conclusion, fully coinciding with those obtained earlier:

``smooth" function $\rho=f(E_{ex})$ reproduces the total gamma-spectrum
of the thermal neutron capture noticeably worse, than the stepped
functional dependencies obtained in [2,3];

in the well deformed dysprosium taking into account various forms of the energy dependence
of the radiative strength functions of primary and secondary transitions substantially
improves the quality of this description.
Moreover the degree of the dependence $k(E1)+k(M1)$ on the energy of the secondary
transition is less than in the primary one. 
In tungsten because of the smaller divergence of the form
of the experimental spectrum and the calculated one similar
calculations are differ less.

The same picture is also observed for the pairs of $\rho$ and $k(E1)+k(M1)$, reproducing
only the  cascade intensities and, simultaneously, the cascade population of levels
below 3-4 MeV.

A more detailed analysis of deviation of  the calculated spectrum from the experimental one
for tungsten shows, that the degree of description of the difference between the
functional dependencies of the radiative strength functions of primary and secondary
gamma-transitions requires more precise definition and detailing.
However, to do this is only possible after obtaining additional information about the
intensities of cascades. Large coefficients of the transfer of the errors of total gamma-spectrum
to the errors of $\delta\rho$ and $\delta(k(E1)+k(M1))$ do not make it possible
using data  of fig. 3 only.
In particular, taking into account the influence of the structure of the excited
levels on a change in the form of the energy dependence of radiative strength functions
most likely should be carried out up to the neutron binding energy.
One must not exclude the possibility that the radiative
and neutron strength functions also depend on the structure of neutron resonances
at the excitation energies larger, than $B_n$.

\section{Conclusion}

The comparison of the total gamma-spectra for different functional dependencies
of $\rho$ and $k(E1)+k(M1)$ both on the excitation energy of nucleus and on
the energies corresponding to the primary and secondary gamma-transitions
for the thermal neutrons capture in Dy and W with the experimental data was carried out.

It showed that
model predictions of the non interacting Fermi gas in these nuclei give worse
correspondence, than the level density from the procedures [2,3].
This conclusion totally corresponded to the one obtained earlier [12].

Large transfer coefficients of the errors of total gamma-spectra to the errors
$\delta \rho$ and $\delta(k(E1)+k(M1))$ directly follow from the comparison of the
data in Figs. 1,2 and 3.  This circumstance confirms the conclusion [1],
that the measurement of such spectra,  for example in the procedure [8],
requires accuracy on $\sim 2$ orders larger, than in the procedure [2,3].
And it limits the possibilities of the independent checking of different
sets of $\rho$ and $k$, both of 
model determined ones and of experimentally obtained ones.
The use of total gamma-spectra for their testing necessary requires
the comparison of different variants of such data.

And even total reproduction of the experimental total gamma-spectrum by calculation
with a certain set of $\rho$ and $k$ is not the proof of the correspondence of these
values to the real parameters of nucleus.
However, explicit nonconformity is a quite single-valued proof of the presence of
larger or smaller systematic deviation for them with the experimental one.

\begin{flushleft}
{\large\bf References}\end{flushleft}\begin{flushleft}
\begin{tabular}{r@{ }p{5.65in}} 
$[1]$ & A.M.  Sukhovoj, V.A.  Khitrov, JINR preprint E3-2005-196, Dubna, 2005.\\

      & http://www1.jinr.ru/Preprints/Preprints-index.html\\
$[2]$ & A.M.  Sukhovoj, V.A.  Khitrov, Par. and Nucl.,  36(4) (2005) 697.\\
      & http://www1.jinr.ru/Pepan/Pepan-index.html (in Russian)\\
$[3]$ & E.V.  Vasilieva, A.M.  Sukhovoj, V.A.  Khitrov, Phys.  At. Nucl. 
64(2) (2001) 153, nucl-ex/0110017\\
$[4]$ & S.G. Kadmenskij, V.P. Markushev, V.I. Furman, Sov. J. Nucl. Phys. 37 (1983) 165.\\
$[5]$ & P. Axel,  Phys. Rev. 1962. 126. $N^o$ 2. P. 671.\\
$[6]$ & V.G. Soloviev, Phys. of Elementary Particles and Atomic Nuclei, 3(4) (1972) 770.\\
$[7]$ & O.T.  Grudzevich et.al., Sov.  J.  Nucl.  Phys. 53 (1991) 92.\\
      & B.V. Zhuravlev, Bull. Rus. Acad. Sci. Phys. 63 (1999) 123.\\
$[8]$ & G.A.  Bartholomew et al., Advances in nuclear physics 7 (1973) 229.\\
      & A. Schiller et al., Nucl.  Instrum. Methods Phys.  Res. A447 (2000) 498.\\
$[9]$ & Lomachenkov I.A., Furman W.I., JINR, P4-85-466, Dubna, 1985\\
$[10]$ &  O.T.  Grudzevich,  Phys.  At. Nucl. 
 62 (1999) 192.\\
$[11]$ & L.V. Groshev et al., Atlac thermal neutron capture gamma-rays spectra,
 Moscow, 1958.\\
$[12]$ & A.M. Sukhovoj, V.A. Khitrov and  E.P. Grigor'ev,
 INDC(CCP)-432, Vienna 115 (2002).\\
$[13]$ & V.A. Khitrov,  Li Chol, A.M. Sukhovoj, XI International Seminar on Interaction
of Neutrons with Nuclei,  Dubna, 22-25 May 2003,
E3-2004-9, Dubna, 2004, p. 92.\\
 & V.A. Khitrov, A.M. Sukhovoj, Pham Dinh Khang, Vuong Huu Tan, Nguyen Xuan Hai,
XIII International Seminar on Interaction of Neutrons with Nuclei,  Dubna, 22-25 May 2005,
E3-2006-7, Dubna, 2006, p. 41.\\
$[14]$ &http://www-nds.iaea.org/pgaa/egaf.html\\ 
       &G.L. Molnar  et al., App. Rad. Isotop. 53 (2000) 527.\\
$[15]$ &W. Dilg, W. Schantl, H. Vonach, M. Uhl, Nucl. Phys. A217 (1973) 269.\\
\end{tabular} 
\end{flushleft}

\begin{figure}
\begin{center}
\leavevmode
\epsfxsize=15.5cm
\epsfbox{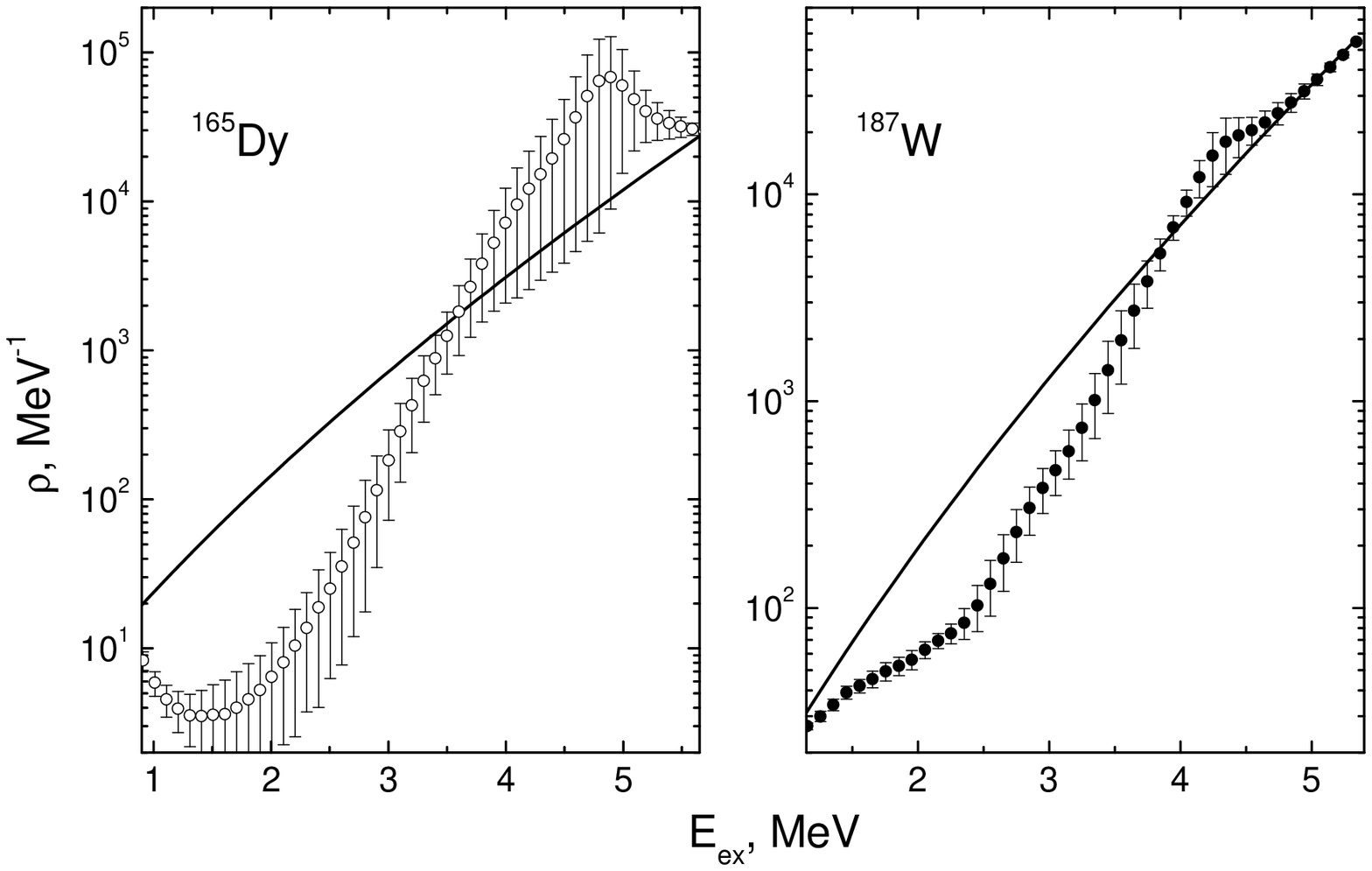}
\end{center}
\hspace{-0.8cm}

\vspace{-12cm}

{\bf Fig.~1.} The interval of probable values of the level density enabling
the reproduction of the experimental cascade intensity (as a function of
primary transition energy) and total radiative
width of capture state. The  line represents predictions of the model [15].
\end{figure}

\begin{figure}
\begin{center}
\leavevmode
\epsfxsize=15.5cm
\epsfbox{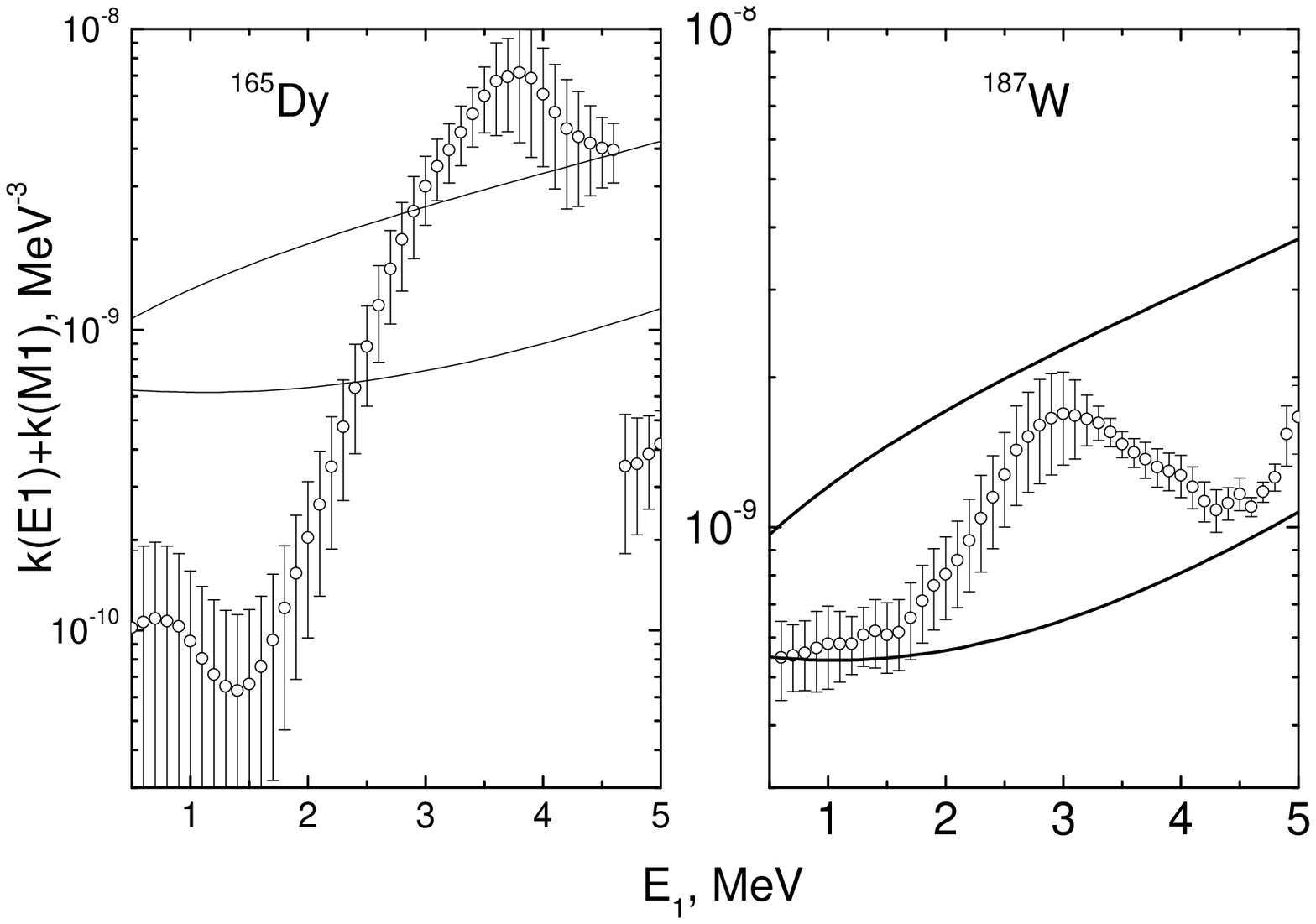}
\end{center}
\hspace{-0.8cm}
\vspace{-12cm}

{\bf Fig.~2.} The probable interval of the sum strength function $k(E1)+k(M1)$
(points with error bars) providing the reproduction of the experimental data.
Solid  line  is predictions of models [4,5].
\end{figure}

\begin{figure}
\begin{center}
\leavevmode
\epsfxsize=15.5cm
\epsfbox{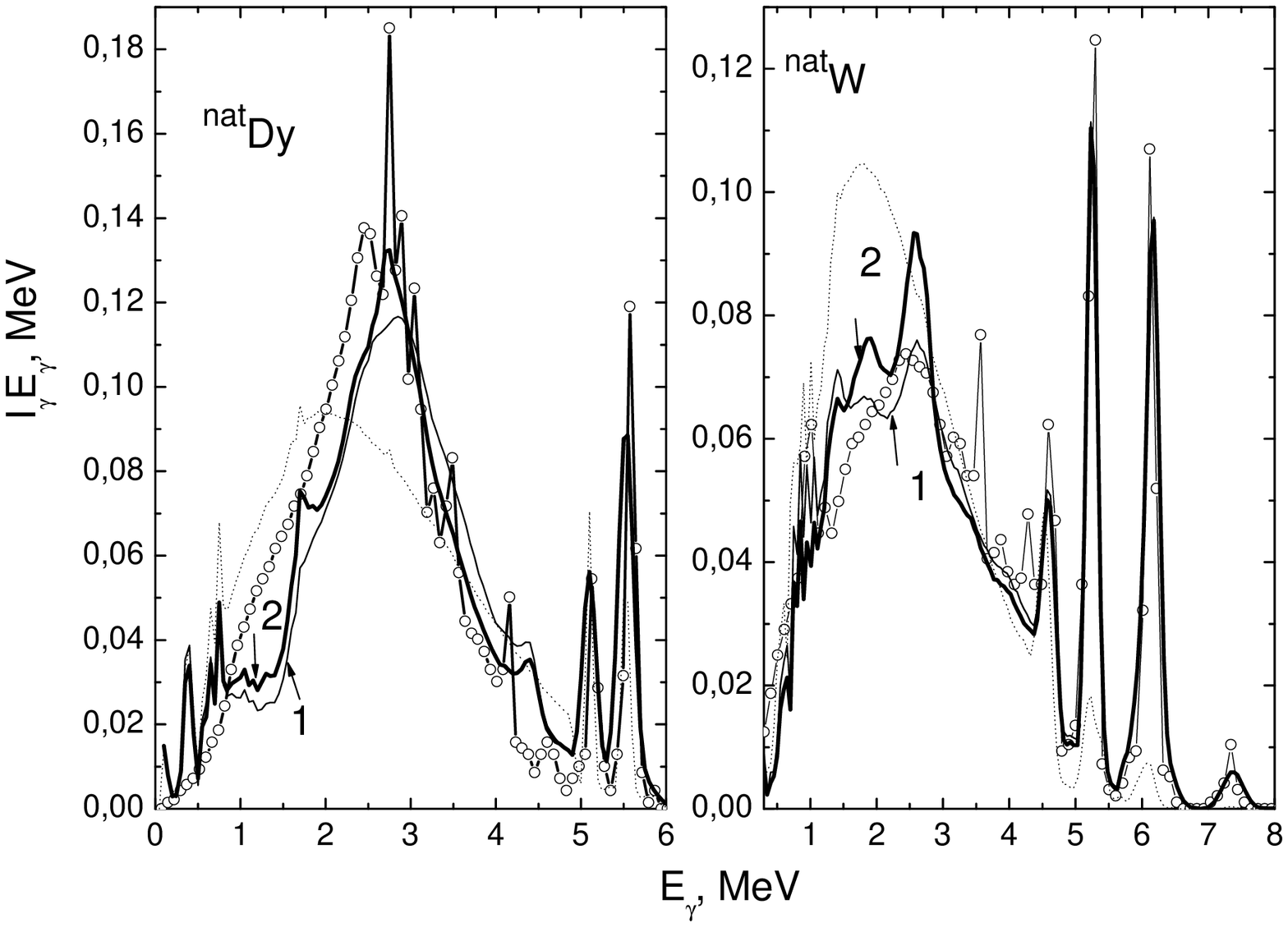}
\end{center}
\hspace{-0.8cm}\vspace{-12cm}

{\bf Fig.~3.} The experimental (points) total spectra of $\gamma$-radiation
following thermal  neutron capture for the
 $^{nat}$Dy  and $^{nat}W$ targets.
 Solid and dashed lines represent results of calculation
using data [2,3] and model parameters [4,5,15], respectively.
Line 1 calculated with
$\rho$ and $k$ parameters from [3], line 2 - from [2], corresponding.
\end{figure}

\end{document}